\documentclass[acmlarge]{acmart}
\makeatletter
\newcommand{\myconfshort}{\acmConference@shortname}
\newcommand{\myconffull}{\acmConference@name}
\newcommand{\myconfdate}{\acmConference@date}
\newcommand{\myconfloc}{\acmConference@venue}
\AtBeginDocument{
  \fancypagestyle{firstpagestyle}{
    \fancyhead{}%
    \fancyfoot[C]{}%
  }
  \fancyhf{}
  \fancyhead[LO]{\@headfootfont\shorttitle}%
  \fancyhead[RE]{\@headfootfont\@shortauthors}%
  \fancyhead[LE]{\@headfootfont\footnotesize \myconfshort, \myconfdate, \myconfloc}%
  \fancyhead[RO]{\@headfootfont\footnotesize \myconfshort, \myconfdate, \myconfloc}%
  \fancyfoot[C]{}%
}
\makeatother
\acmBooktitle{\conffull\@ (\confshort), \confdate, \confloc}

\AtBeginDocument{%
  }

\copyrightyear{2026}
\acmYear{2026}
\setcopyright{cc}
\setcctype{by-nc-nd}
\acmConference[FAccT '26]{The 2026 ACM Conference on Fairness, Accountability, and Transparency}{June 25--28, 2026}{Montreal, QC, Canada}
\acmBooktitle{The 2026 ACM Conference on Fairness, Accountability, and Transparency (FAccT '26), June 25--28, 2026, Montreal, QC, Canada}
\acmDOI{10.1145/3805689.3806739}
\acmISBN{979-8-4007-2596-8/2026/06}

\begin{document}

\title[]{Reckoning with the Political Economy of AI: Avoiding Decoys in Pursuit of Accountability }


\author{Janet Vertesi}
\affiliation{
\department{Sociology Department}
\institution{Princeton University}
\country{Princeton, NJ USA}
}
\email{jvertesi@princeton.edu}

\author{danah boyd}
\affiliation{
\department{Department of Communication}
\institution{Cornell University}
\country{Ithaca, NY, USA}
}
\email{dmb478@cornell.edu}

\author{Alex Taylor}
\affiliation{
\department{Institute of Design Informatics}
\institution{University of Edinburgh}
\country{Edinburgh, UK}
}
\email{alex_taylor@acm.org}

\author{Benjamin Shestakofsky}
\affiliation{
\department{Department of Information Science}
\institution{Cornell University}
\country{Ithaca, NY USA}
}
\email{bs889@cornell.edu}


\begin{abstract}
The Project of AI is a world-building endeavor, wherein those who fund and develop AI systems both operate through and seek to sustain networks of power and wealth. As they expand their access to resources and configure our sociotechnical conditions, they benefit from the ways in which a suite of decoys animate scholars, critics, policymakers, journalists, and the public into co-constructing industry-empowering AI futures. Regardless of who constructs or nurtures them, these decoys often create the illusion of accountability while both masking the emerging political economies that the Project of AI has set into motion, and also contributing to the network-making power that is at the heart of the Project's extraction and exploitation. Drawing on literature at the intersection of communication, science and technology studies, and economic sociology, we examine how the Project of AI is constructed. We then explore five decoys that  seemingly critique\textemdash but in actuality co-constitute\textemdash AI's emergent power relations and material political economy. We argue that advancing meaningful fairness or accountability in AI requires: 1) recognizing when and how decoys serve as a distraction, and 2) grappling directly with the material political economy of the Project of AI. Doing so will enable us to attend to the networks of power that make ``AI'' possible, spurring new visions for how to realize a more just technologically entangled world.
\end{abstract}



\maketitle

\section{Introduction}
The \textit{Project of AI} is one of world-building, but its activities stretch far beyond the development of new technologies aimed at advancing societal progress. Rather, those who are constructing AI\textemdash including wealthy financiers and highly capitalized corporations \textemdash are actively reinforcing particular networks of power and wealth while unseating others. The technical systems that they create are co-constituted alongside this world-building effort. 

In this paper, we draw attention to the particular arrangements and configurations of novel networks that constitute the material political economy of AI: a system that enrolls infrastructure, financing, and relationships among powerful stakeholders, all the while creating the impression of competition, inclusion, and governability. As part of their world-building endeavor, those constructing what we herein call the Project of AI implicitly encourage and explicitly direct researchers, technologists, policymakers, and critics from civil society to focus on the technical particulars associated with ``AI,'' including its potential uses, future capabilities, and the extent to which the foundation models are inclusive or biased. We posit that the result is a set of decoys that shift attention away from the agglomeration of world-making actors, their entangled relationships, and alternative framings of our sociotechnical future. This type of misdirection\textemdash intentional or not\textemdash undermines efforts to fully appraise what the Project of AI is, how it is working, and where the levers of power and accountability are located.

The decoys we offer are not incidental to the Project: they are integral to its co-construction. Through cultural mechanisms and seemingly irrefutable, familiar logics, decoys provide the rationale for AI and its particular configuration of political actors and economic structures. In other words, the misdirections enabled by decoys uphold and co-construct the political economy of AI. Moreover, these decoys enroll outsiders into the Project of AI by orienting their attention to its particulars. This diverts potential challengers from tracking the locus of accountability and limits their ability to launch the forms of critique\textemdash social, technical, or otherwise\textemdash necessary to ensure transparency and fairness. We argue that holding the right actors to account requires that we expand our focus to interrogate the political economy of AI. In this paper, we start this work by examining the conditions set by these decoys. 

We first lay out the theories of network power that guide our analysis of the Project of AI, including assembly mechanisms for its material political economy. We then explore five decoys to illustrate how each appears to offer a tractable way forward for accountability and governance, but in fact conscripts us into participating in the Project of AI itself. Each decoy, we argue, further enables AI brokers' expansion of wealth and power, the architecture of flows, and the conditions for continued growth. Perversely, AI companies and their financiers benefit from the efforts of scholars, journalists, critics, and lay publics, because our collective attention to decoys upholds, stabilizes, and reinforces the Project of AI. This has pressing implications for how we articulate and intervene in pursuit of fairness, accountability, and transparency.

\section{Theories of Network Power in Political Economy}
Much as sociotechnical analysis requires grasping how technologies and social settings are entangled, a political economy analysis interrogates the relationship between the complex and entangled forces of politics, markets, and society. As a social science approach, its focus is on how social, political, and economic forces interact to shape the material conditions of possibility. Diverse actors and the minutiae of shifting relationships, laws, and labor link up to form an agglomeration of agents and structures that exert continuing power and influence. Network-scale thinking is needed to observe how the capabilities for action shift with the accumulation of power and materials, rather than assuming static nodes or stable connections.\footnote{Our perspective on the concretization of power in technological systems draws on numerous intersecting theories of networks. Even if the scholars we draw on may use ``network'' to refer to somewhat disparate phenomena, together they point to ways in which powerful players seek to rearrange and configure material, social, and economic orderings to their benefit.} 

Using this thinking as a prism to examine the Project of AI, we draw upon scholarship from communication, science and technology studies, and economic sociology to show how actors assemble the material infrastructures of computing and financialization that together form the foundations for extraordinary degrees of authority and power. In brief, heterogeneous capitalist actors have seized the moment offered by a period of uncertainty to mobilize communications technologies and financial relationships to their advantage. In doing so they are re-structuring markets in their favor, and directing the flow and capture of capital, resources, data, materials, and labor between sites. 

A handful of companies, countries, and elites are battling each other to realize AI into being \cite{Srnicek_2026}. Many of the better-known architects of the Project of AI are established players in Silicon Valley, emerging from existing networks (such as the `PayPal Mafia' and YCombinator) and interweaving established companies (such as Nvidia, Intel, or Microsoft). As communication scholar Manuel Castells observed in the world-making project of Rupert Murdoch's media empire in the 1990s and 2000s \cite{Castells2009,Arsenault_Castells_2008}, novel information technology and communication architectures present opportunities for the consolidation of power among networked elites, which he calls \textit{networks of power}. ``Networks'' have a dual meaning for Castells, as they are simultaneously social networks (i.e., relations among elites in the media landscape) and communication networks along which broadcast signals can flow. He observed how elite ``programmers'' and ``switchers'' configure networks in their favor. ``Network-making power,'' according to Castells \cite{Castells2009,Castells2011}, is the highest degree of power in an information society, as advantaged individuals, their associates, and corporations can use their position, possessions, and economic influence to achieve their goals and conscript others into their shared project. The concatenation of network power produces implications for governance and economic capture, as the information societies that result are governed not by codes of conduct or regulatory oversight but by the relational ties and moral codes \cite{Morrill1991} of those who hold the power to program and switch the network in their favor. The parallels between Castells' broadcast networks and today's AI sector are striking. 

Castells' heterogeneous networks recall early work from Actor-Network Theory in science and technology studies (STS) which attends to new forms of agency that emerge as humans and non-humans (technologies, intelligences, institutions, and materials) assemble in a network \cite{Latour1990,Law1987}. More recent work in STS on financial market-making addresses the agglomeration of power and capital under what appear to be external ``laws'' for action and interaction, like ``the free market.'' In reality, these laws are not naturally occurring but purposefully built into the assemblage from the ground up \cite{MacKenzie2021,Callon2007,MuniesaCallon2007,MacKenzie2006}.  A ``material political economy'' \cite{MacKenzie2017} expresses how these political and economic arrangements of institutions and economic exchange are locally grounded and physically instantiated. We see this in the Project of AI as the weighty new, asset-heavy materialities of infrastructure and labor that make these arrangements ``durable'' \cite{Latour1990}. For example, data centers and off-shored farms fuel large-scale LLMs while embedding power relations into their very pipelines \cite{Srnicek_2026, Star1999}.

The market context for the Project of AI is therefore both found and made. It is found, in that it emerges in the wake of considerable uncertainty, social change, and the retreat of democratic institutions. Economic sociologists argue that moments of uncertainty offer key entry points for novel players to take hold of the narrative \cite{beckert2018,Schumpeter_2008}. The emergent constellation of people, organizations, and financial structures that are currently configuring ``AI'' as we know it were already ascendant in Silicon Valley in the aftermath of the dot-com bubble. They grew in economic and political power following the Great Recession of 2008, as the banking sector on Wall Street increasingly joined forces with elites in Silicon Valley \cite{MorrisSullivan2020}. And unlike other sectors in the economy that suffered under Covid-19, the pandemic period disproportionately benefited both tech companies and the financiers who depend upon and enable the tech industry \cite{Gain2022}. By 2022, the tech industry was flush with cash, with investors ravenously hunting for the next unicorn. Efforts to manifest crypto and the metaverse had failed. OpenAI's public release of ChatGPT in December 2022 created the opening for market-making that many in these elite networks were seeking.

The market for AI is also made, in the sense that markets are not subject to natural laws but emerge from constellations of social agents (i.e. they are ``embedded'' in social worlds \cite{granovetter_economic_1985}). As firms seek to convince (or deploy their power to compel) regulators, competitors, and customers to go along with their vision, they contribute to determining how the market should be structured and how capital should flow \cite{Fligstein1996}. Moments of market emergence are particularly fluid times marked by the founding of new firms, the entrance of existing firms, price wars, and internal power struggles within organizations. Markets stabilize once competitors, suppliers, and customers can constitute and shape durable networks that offer economic actors information about each other’s behaviors, opportunities to form collective identities, and shared understandings of how the game is played. Well-connected firms are better positioned to navigate uncertainty, cultivate legitimacy, and secure the customers and suppliers needed for survival \cite{FligsteinCalder2015}. Dominant firms can secure their position by influencing public policy, pushing states to remove regulations, offer subsidies, enforce property rights, or create new rules that impose prohibitive costs on challengers who wish to enter the market. In this way, even as dominant players compete with each other, they simultaneously create a stable order in which the most advantaged firms can reliably sustain and grow profits \cite{Fligstein2002}.

Although many of the features of our contemporary moment can be found in prior technology bubbles \cite{GoldfarbKirsch2019,Saxenian2000}, bringing these ideas about networks, materials, and market mechanisms together gives us fresh insight into the Project of AI. Across the sector's companies, famous faces, and machineries (whom together we label ``AI brokers'') we see the emerging configuration of these non-technical elements of network power as one of the primary characteristics defining the age of AI. Technologies like LLMs, chatbots, and image generators have varying degrees of capability, but their very condition of possibility emerges from the manipulation of materials, ideas, and capital. 

In a typical history of technologies, markets, and capitalism, such activities take place under the eye of some form of regulatory state. However, as tech oligarchs leverage their infrastructural and political power to reconstruct the state in their own image \cite{Cohen2025}, today's market-making brokers engaged in the Project of AI enjoy little regulatory oversight, placing few limits on the rush to establish an elite cadre of dominant individuals, corporations, financiers, and infrastructures under the aegis of building ``AI.'' Meanwhile, the second Trump administration has advanced policies supporting the build-out of data center and energy infrastructures domestically while pushing for the export of U.S. AI technologies abroad \cite{Chen2026}. Tech titans are currently consolidating control over every aspect of the AI supply chain\textemdash including energy inputs, chips, foundation models, compute, and software development tools\textemdash to ensure that their firms remain central players as AI grows \cite{vanderVlist2024}. Major cloud computing providers are foregoing the arms-length, contractual inter-firm relations that previously prevailed in the tech industry, choosing instead to partner with, invest in, and acquire multiple companies at each layer of the AI stack. Microsoft's partnership with OpenAI is one example of a movement toward inter-firm collaboration for mutual benefit, even as the two companies function in practice as ``frenemies'' \cite{TanThelen}.

Like all heterogeneous networks, these networks of power can be contested or destabilized as they are under assembly and can even be called to account. However, doing so requires taking aim at the network and market-making enterprise, instead of at individual players or product specifications.

\section{Decoys for the World-Making Project of AI}



Many existing critiques of AI and efforts to hold actors to account have been concerned not with the reach and construction of network power and its corresponding political economy, but with decoys. In deploying the term ``decoy,'' we allude to the category of activities that serve the Project of AI through misdirection. Some decoys are purposefully constructed, but others have emerged more organically, often created by well-intended stakeholders who imagine that they are challenging AI. When decoys draw attention to technical details or even legal specifics, this leaves the underlying assembly of network power unexamined. 

Decoys can best be understood as a manifestation of misdirection. Magicians use misdirection as a form of cognitive deception to create an illusion. The decoys magicians employ in such trickery help to hide the mechanisms of the trick itself. Other kinds of decoys are meant to serve as traps. For example, hunters use decoys to lure vulnerable prey into a trap even as wildlife conservationists use decoys to attract birds to safe places (i.e. away from airports). Other decoys are simply a way to orient or control attention.  In each of these cases, decoys are not just a distraction from an overall endeavor; they are also a crucial tool for configuring an environment. Each decoy allows network-making power to recede into the background or even to go entirely unnoticed\textemdash our attention is drawn to the decoy even as structures are put in place that allow the trick to unfold. While we focus on the technical specifics of AI, brokers establish flows to expand their wealth and power. Decoys also enroll others to actively assist in the co-construction of the Project of AI, its network of power, the scope of its associated market, and the overall cultural myth-making essential to stabilizing a market around AI. In this way, even critics build momentum toward the Project of AI. Decoys therefore form accountability traps that, perversely, enable rather than constrain the powerful network assembly behind the Project of AI.

The Project of AI is flush with rhetorical and material decoys which, like myths, are ``captivating fictions'' notable for what they both reveal and conceal \cite{Mosco2004}. Some are purposefully constructed or leveraged by AI brokers to draw attention toward specific aspects of the Project of AI (and away from others). Other decoys have emerged in the economic and political scrum as actors seek to understand and control the future of AI. Regardless of how they came into being, the decoys do provide a means to understand the Project of AI and its political economy or prospective governability \cite{foucault2004security}, although not if we take them at face value. Instead, reading the decoys as intellectual and material traps can reveal how certain forms of thought and knowledge\textemdash and the (infra)structures enabling them\textemdash create the conditions for the construction and concentration of network power. 

To highlight the work that decoys do in co-constructing the Project of AI and organizing power\textemdash and to help to inform more robust critical analyses\textemdash we examine five distinct but entangled decoys: the ontological decoy, the inevitability decoy, the disruption decoy, the safety decoy, and the regulatory decoy. As we show, each decoy draws critics' attention away from the Project of AI while contributing to its very architecture at the same time. Learning to see these configurations is one way to prevent them from functioning as traps.

\subsection{The Ontological Decoy}

Anthropologist Lucy Suchman \cite{Suchman2023} notes that ``AI'' is simultaneously posited as ``a thing'' with distinct properties in the world and a ``floating signifier'' that ``works to escape definition in order to maximize its suggestive power.'' This slipperiness can be powerful because it often entices different actors to obsess over how to bound what AI is or should be, rather than to focus on the work the Project of AI is doing in the world. Yet there can be no resolution to a debate about what AI is or is not, what it can or cannot do, or what it conceivably could or should do or be in future. That is not because AI is inherently ambiguous, but because maintaining the conceptual ambiguity of AI serves the interests of AI brokers. The \textit{ontological decoy} shifts the terms of the debate to \textit{what AI is}, drawing attention away from \textit{the work that AI is doing}: that is, enabling the expansion of the Project of AI. While seeking ontological clarity about AI in the face of strategic deployments of ambiguity might seem like a check to power, it actually reinforces the centrality of this slippery definition of AI.  

Consider changes to the funding ecosystem that arose in 2023. Countless entrepreneurs seeking venture capital, academics seeking grants, and civil society organizations seeking philanthropic support suddenly found themselves forced to spin their activities as contributing to\textemdash or, in some cases, being critical of\textemdash the advancement of AI to successfully obtain capital. To do so, many had to treat ``AI'' as an ambiguous term to justify the relevance of their work. Of course, they were not alone. Those doling out money\textemdash including philanthropic organizations, government agencies, and venture capitalists\textemdash felt pressure to shift funding strategies in response to the ``AI moment.'' Fearing being left behind, many others followed suit \cite{DiMaggioPowell1983}. Those seeking and distributing capital leveraged definitional ambiguity, even as scholars, journalists, and politicians fretted over how to bound AI. Pursuing ontological clarity does nothing to challenge the rearrangement of financial flows. 

When powerful networks of actors strategically leverage ontological ambiguity to assert authority and grab power, what emerges is not a coherent or bounded technological object. Instead, such rhetorical flexibility feeds an accretion disk of people, ideas, materials, and momentum that forms and informs the emergent marketplace. On the one hand, these become the materials of capitalist ``creative destruction'' \cite{Schumpeter_2008}. On the other, firms engage in acts of cultural entrepreneurship, attempting to demonstrate an alignment between their commercial interests and public morals in order to better develop market fit \cite{Zelizer1985}. Answers to the question of ``what AI should be'' are thus part of market-making processes. 

Many institutions interact with this question in ways that, wittingly or not, support a broader cultural project. For instance, the business practices and dispositions surrounding the ``new economy'' in the late 20th century were taken up and disseminated by knowledge producers like academics, management consultants and gurus, managers themselves, media, and government agencies \cite{Thrift2001}. This swirl of actors constituted a ``cultural circuit of capital'' whose discourses helped bring about an ``imagined future'' of asset value inflation by performing it into being during the dot-com boom \cite{Beckert2016}.

This is how the ontological decoy constitutes a form of entrapment. It continually draws actors into unanswerable questions around determining and clarifying the nature of AI, even as we observe how the peeling back of technical specifics fails to render these matters more or less certain \cite{Buolamwini2023,BenderGebruMcMillan-MajorShmitchell2021}. Maintaining the fluidity of ``AI'' and its possibilities across a heterogeneous coalition of materials, funders, libraries, and people keeps the decoy active. Instead of certainty, we observe instead a form of ``ontological gerrymandering'' among participants \cite{WoolgarPawluch1985}, where members of the cultural circuit of capital choose to focus on one aspect of AI at the expense of another, repeatedly changing the frame. Meanwhile, public talk about AI and its purported capabilities now or in the future supports the perpetuation of the ontological decoy, even when the discussion appears to divide critics into different ideological camps\textemdash such as those who wish ``it'' to be controlled by ``the good guys,'' those who wish to see a ``nationalist'' Project of AI, or those who believe it will usher in a post-human and potentially off-Earth future \cite{BohnerVertesi2025}.  

We see this clearly in the configuration of actors around AI at the time of writing. Anthropic, OpenAI, Google, and others are currently operating in a competition governed not by an arm’s length open market, but by economic conditions articulated and rehearsed by tech leaders, investors, lenders, and corporations that stand to benefit from the language of competition, growth, and one-upmanship \cite{Fligstein2002}. Even the narrative of competition and differentiation benefits this network, as it enables companies to push their workers to move faster and work harder, driving market values higher. Meanwhile, continued ontological instability around AI allows brokers to push AI into ever more sectors and practices. Smaller companies are labeling their technologies ``AI'' to capture financial resources and obtain network power. Such activities appear to demand that we ask, ``is this really AI?'' or even ``is this a \textit{good} use of AI?''  

Rather than getting trapped by discussions about ``what AI is and how AI should work,'' we would do better to expand our scope to understand how ``AI'' is taking all the air out of the room. Doing so allows us to see the unfurling heterogeneities and shifting networks involved in the Project of AI\textemdash and to interrogate why brokers maintain strategic ambiguity to ensure that ``AI'' is flexible enough to adapt as their market-making unfurls. We can observe how product differentiations or evolving foundation models shift along with unstable economic configurations, even when these changes appear to respond to internal or external benchmarks. We must therefore resist the ontological decoy and its imperative to specify AI's ``true'' nature or potential. Effective critique must destabilize network power instead of feeding the circuit.
\subsection{The Inevitability Decoy}

The tech industry often uses the rhetoric of inevitability to justify its developments \cite{CheneyLippold2025,LeonardiJackson2004}. Thus it should be unsurprising that the language of inevitability accompanies the hype cycle surrounding AI \cite{BareisKatzenbach2022,Markelius2024,bender_ai_2025}. Yet what makes this language notable in the context of AI today is the success that AI brokers have had in enrolling others to embrace this rhetoric. The \textit{inevitability decoy} has been deployed by a wide array of interlocutors to construct a shared project of market capture. Cultural narratives about AI therefore often conceal the material power moves that the story of inevitability enables \cite{Mosco2004}. 

One source of the inevitability decoy's utility is how it enables actors to repeatedly shift temporalities, offering new projections as to when the future promise of AI will ultimately be enacted. Whether this future concerns the arrival of AGI, quantum computing, or other giant leaps in capability, AI brokers benefit from discussions that usher the event horizon closer or further away while maintaining the frame of inevitability \cite{BrennanKakWest2025}. This discursive work is particularly important for entrepreneurs and financiers who, in their efforts to acquire resources or create or capture a market, must wield articulations of the future \cite{beckert2018}. Thus, inevitability rhetoric mobilizes acolytes and resources toward a projected, future technology of which the current instantiation is typically only the beginning \cite{MesseriVertesi2015, TavoryEliasoph2013}. At the same time, when powerful actors project a story of technological determinism and inevitability, they enhance their own economic and social capital and constrain other actors and futures in the process \cite{GoldfarbKirsch2019}. This lays the groundwork for network capture, even monopoly \cite{Narayan2023,Srnicek2016,VertesiGoldsteinEnriquezLiuMiller2020}. 

Anticipatory future-making also naturalizes AI's phased stages of development and deployment, such as the roll-out of data centers. Even the contingencies of local setbacks are staged within a larger narrative that, usefully, requires investment today to circumvent roadblocks to the purportedly inevitable future \cite{beckert2018,MesseriVertesi2015}. Inevitability rhetoric also normalizes various types of risk, including economic and technological risk, because when the future is pre-determined, risky business decisions are reconstituted as necessary. Inevitability talk therefore offers a form of ``discursive closure'' that can rush certain futures in and prevent other futures from taking hold \cite{LeonardiJackson2004,Markham_2021}. 

 The sociology of religion reveals how such projected futures do not need to come true to ensure followers' deep commitment. To the contrary, a shared orientation toward a promised future\textemdash like ``the second coming'' or an afterlife\textemdash can be effective in enrolling people to the cause precisely because it is unspecific in its details and can speak instead to inchoate, deeply held beliefs \cite{Blili-HamelinHancox-LiSmart2024,Friedland1964}. Thus acolytes believe that with enough talent and capital, the future they are promised actually \textit{is} inevitable, and is only blocked by a lack of resources in its path.

The repetition of similar futuristic claims across numerous companies constructs the appearance of coherence, suggesting that all involved in the AI Project have clarity about AI's role in ``our'' future \cite{BareisKatzenbach2022}. Meanwhile, underlying competition and collaboration among central players performs this inevitability into being through the assembly of chips, data centers, software and data conduits. In the process, AI brokers create a set of conditions that make AI ``too big to fail'' even as individual companies come or go \cite{BrennanKakWest2025}. Continued discussion of AI's inevitability therefore contributes to resource acquisition and market manipulation. By centering particular myths about the market, it sets the conditions for who controls, owns and exerts influence in the Project of AI.

The purported race between major world powers to build AGI contributes to inevitability discourse. Proponents of AI invoke a mixture of historical examples of technological competition, especially WWII's Manhattan Project or the Cold War space race. These historical fictions conveniently leave out nuclear scientists' guilt and subsequent dedication to non-proliferation \cite{Wellerstein2026}, the resources and global hierarchies subsumed in nuclear-era diplomacy \cite{Hamblin2021,Hecht2014,Wolfe2013}, and distrust in public institutions following accidents \cite{Sagan1995,Downer2014,Vaughan1996}. Instead, AI brokers tell these stories to focus on the need for the ``good guys'' to prevail in constructing the singular pathway that guarantees the future\textemdash controlled by a small set of actors with the unique power to enact that promise. The ``AI is inevitable'' narrative thereby upholds asociotechnical imaginary that imbricates state and corporate power with shared narratives about promised futures \cite{JasanoffKim2015}. Such narratives are attractive in no small part because they suggest the need for material support from governments at federal, state, and local levels, while suppressing concerns that could impede the achievement of this purportedly necessary good. The combination of geopolitical considerations and ``inevitability'' discourse also helps to justify political and economic commitments, including continued antagonism between the U.S. and China, and strategic investments by hegemonic powers in countries whose natural resources can fuel the Project of AI.

Ironically, the inevitability decoy extends to efforts to engage on-the-ground stakeholders through participatory design \cite{sloane_participation_2022,Sloane_2024}. Engaging potentially impacted communities is seen as a normative good, but such interventions enroll widening swaths of people into realizing this ``inevitable'' future. As ritual, stakeholder participation upholds inevitability by articulating apparently important components of this eventual promise \cite{Ames2019}. Such work therefore feeds rather than resists the Project of AI. These efforts underscore how the concerted attention of well-meaning people can ameliorate the negative consequences of AI's inevitable march by shaping it appropriately (i.e., reducing bias). They also shore up AI brokers' influence, leaving decisions that reinforce their networks of power\textemdash including the ends to which capital is directed, which resources are extracted, and how and markets are made\textemdash untouched.

\subsection{The Disruption Decoy}
Shortly before Facebook's 2012 IPO, when Mark Zuckerberg filed the company's prospectus with the Securities and Exchange Commission, he outlined ``five core values'' that shaped how executives ran the company, including their local saying: ``move fast and break things'' \cite{Ebersman2012}. This phrase has since become a motto adopted across the tech industry, reinforcing the widespread misinterpretation \cite{ChristensenRaynorMcDonald2015} of Clayton Christensen’s theory of ``disruptive innovation'' \cite{Christensen1997}, which many have interpreted as unequivocally celebrating all forms of market disruption as inherently constructive.

It is thus perhaps unsurprising that AI is hailed as a ``disruptive'' technology in Silicon Valley. But just what AI disrupts\textemdash and the true value of that disruption\textemdash is slippery. One dominant narrative is that AI will disrupt the labor market by replacing a wide swath of menial tasks. Tech leaders celebrate this disruption, promising more efficient businesses even as they publicly raise concerns about the potential for large-scale job loss. In doing so, they reinforce their belief that innovation should be pursued without regard for its social consequences. But as tech leaders, academics, and pundits debate how much job loss will occur because of AI\textemdash and how AI will transform the workforce more broadly\textemdash they obfuscate the disruptive move that the brokers backing the Project of AI seek to make. The \textit{disruption decoy} points to this effect: one of highlighting local disruptions as a form of business-as-usual (``efficiency-seeking,'' for instance), while architecting massive changes in the concentration of power across industrial sectors, even across national boundaries.

``Disruption'' carries significant cultural weight. Yet myriad studies show that novel technologies do not disrupt existing social orders and inequalities: they rather serve to reify or entrench existing interests \cite{winner_artifacts_1986,Benjamin_2019,barley_technology_1986,Kling_1991,MacKenzie_Wajcman_1999}. More usefully, in economic sociology ``disruption'' speaks to a market scenario in which both entrants and incumbents seize on moments of uncertainty to secure a market order that favors their interests \cite{Uzzi1999,Fligstein1996}. It is in this latter sense of ``disruption,'' not the former, that we observe how those inhabiting powerful network positions\textemdash in this case, AI brokers\textemdash are well positioned to obtain more money and/or power by leveraging unsettled times to their advantage.

The outcome of such disruptions is never predetermined, yet the power of these networks is that they can shape disruptions in a manner that increases the likelihood of specific outcomes. We can trace these weighty transfers and flows of power through evolving processes of valuation that accompany network consolidation \cite{Elder-Vass2021}. For instance, the accumulation of market value among the ``Magnificent Seven'' \cite{Grobys2025} may serve less as an indicator of these companies' technical potential, speaking more to their leaders' ability to accrue both economic and social capital while positioning their firms as central to a densely connected network of participants \cite{FligsteinCalder2015}. Such an arrangement and its resulting consolidation of social power is visible in the circulation of currencies such as data centers, chips, and reciprocal investments \cite{ForgashGhosh2025}. In other words, the stock market valuation of these companies balloons as an indicator of the concentration of power among an elite few who are well positioned to call the shots, rather than representing the use value of a technology and its ``disruptive'' capacity to achieve its promise.

Despite the tech industry's valorization of ``disruption,'' AI brokers do not like to \textit{be} disrupted. When the Chinese company DeepSeek launched its first model in 2025, it challenged both the dominance of U.S. companies and their own business models, which were premised on AI's high cost and data requirements. Rather than absorb DeepSeek into the  Project of AI, many AI brokers in the United States engaged in gatekeeping to shut out DeepSeek from the broader Project.

AI brokers also use ``disruption'' to direct attention away from their efforts to reorganize firms and hoard capital flows. For decades, financialized firms have sought to reinvent themselves through successive reorganizations that accrued increasing ``rents'' to the company's owners\textemdash for instance, through the minimization of middle management \cite{vertesi_ghost_2025} or the rise of stock buybacks \cite{PalladinoLazonick2024}. AI companies continue this trend by organizing their distribution of functions among global and outsourced outposts \cite{Tubaro2021}. It is certainly true that individual workers' tasks are altered by the introduction of AI, but it is also the case that companies across sectors are using the premise of introducing AI to their workforces as a decoy for run-of-the-mill corporate restructuring \cite{Rajgopal2025}. Meanwhile, firms move core tasks to low-cost talent centers, where automated tools aim to increase the productivity of distant and untraceable workers \cite{KelloggValentineChristin2020}. Citing ``AI'' as ``disruptive'' force provides a justification for both layoffs that will require remaining workers to ``do more with less,'' and the movement of entire sectors of the workforce offshore into alternative regulatory environments far from the purview of federal labor statistics or government oversight \cite{Adams-Prassl_Abraha_Kelly-Lyth_Silberman_Rakshita_2023}. Although research into AI adoption and use in workplaces is doubtless important, such a perspective is also incomplete. When investors' interests compel firms to reorganize massive workforces across multiple sectors and jurisdictions in order to seek new efficiencies and evade accountability, attention to the micro-practices of work with an AI tool is much like cataloging the proverbial deck chairs on the Titanic.

Although ``disruption'' is wielded as a decoy, we believe that AI does in fact represent a significant disruption to work, if not in the popularly understood sense. Far more disruptive than workplace reskilling \cite{braverman_labor_1974} is the renewed and concentrated dependence upon a distribution of labor into invisible and unaccountable zones \cite{GraySuri2019,Tsing2005}; and far more disruptive to politics as usual is how this arrangement supports the extraordinary consolidation of power among a networked elite. The Project of AI commands and conceals a profound infrastructural disruption taking place with respect to the circulation of capital, the reorganization and distribution of global labor, and the concentration of meaningful materials (data, chips, data centers) among a small set of powerful actors. By inviting the public to debate how AI might disrupt the workforce or the progress AI could bring, the disruption decoy distracts us from seeing and debating AI brokers' underlying network-making moves.
\subsection{The Safety Decoy}
Although ``AI Safety'' is front and center for many involved in the Project of AI, the slipperiness of this concept also gives rise to another decoy. In both scholarly circles and public parlance, ``AI safety'' refers to ensuring that AI systems are reliable, do not cause harm, and are designed to reflect broader social values \cite{AhmedEtAl2024,BecerraEtAl2025}. Feminist data critics go further to force a reckoning over sustainability, power, and pluralism \cite{KleinDIgnazio2024}. Safe and responsible AI communities often emphasize commitments like fairness, accountability, and transparency; technical practitioners also point to concepts like ``alignment.'' However, many AI brokers yoke the language of ``AI Safety'' to existential concerns. Harking back to computer scientist Eliezer Yudkowsky \cite{YudkowskySoares2025} and philosopher Nick Bostrom \cite{Bostrom2017}'s arguments at the turn of the millennium, they depict AI's (inevitable) progression to Artificial General Intelligence (AGI) as posing an existential risk to humanity. In a throwback to the inevitability decoy, the brokers of the Project of AI therefore declare it imperative to invest in ``AI safety'' to prevent ``superintelligent'' AIs from destroying humanity, in what is pejoratively labeled ``the Terminator scenario.''

The \textit{safety decoy} refers to conversations and activities that engage with questions of AI safety as questions about contextual use and technical details, while bracketing the consequences of the reproduction of the Project's network of power. Invocations of existential risk are often used to promote tools that massively increase inequality, cause environmental damage, and harm already vulnerable populations under the guise that AI companies are the actors best positioned to mitigate existential risk \cite{beck_risk_1992}. In justifying its approach to safety preparedness, OpenAI argues, ``Some people in the AI field think the risks of AGI (and successor systems) are fictitious; we would be delighted if they turn out to be right, but we are going to operate as if these risks are existential'' \cite{OpenAI2023}.  

Critics are increasingly pushing back against this perspective. After all, this orientation towards safety is disconnected from broader ethical considerations \cite{Nissenbaum_2024}. In calling AI a ``normal technology,'' computer scientists Arvind Narayanan and Sayash Kapoor emphasize how superintelligence has become a distraction from the very real problems that emerge in an era of AI. As scholars of technological risk assessment have noted, AI safety efforts aimed at reducing ``general'' harms are unlikely to succeed given that the risks posed by new technologies are typically unequally distributed across societies and social categories \cite{giddens_risk_1999,beck_risk_1992}. The task of adjudicating the mitigation of AI's ``existential'' risk has largely been left to the actors who will suffer least from its unfurling consequences. In this way, the existential risk approach to safety obscures a range of ideologies that animate many AI brokers and underlie the Project of AI. Computer scientist Timnit Gebru and philosopher Émile P. Torres call attention to these lines of thought under the neologism TESCREAL (Transhumanism, Extropianism, Singularitarianism, Cosmism, Rationalists, Effective Altruism, and Longtermism) \cite{GebruTorres2024}. Among those who embrace these ideologies, AGI is framed not only as inevitable, but also as posing an existential risk to humanity, requiring ongoing efforts to make AI safe. 

As powerful actors' ideological commitments shape discourses around AI, then, ``safety'' has evolved from a meaningful\textemdash if conflicted\textemdash framework into a form that intertwines the ontological, inevitability, and regulatory decoys. Companies now use the language of ``safety'' to signal different things to different communities. In some cases, they deploy the rhetoric of safety to investors in an attempt to differentiate their products and services from other offerings. For instance, most of the leaders of Anthropic left OpenAI over concerns that the organization was not sufficiently addressing existential risk concerns. Anthropic has described its ``portfolio approach to AI safety'' as including both the commitments of those attached to the responsible AI framework, and those who viewed AGI as an existential risk \cite{Anthropic2023}. In early 2026, Anthropic also made much of its resistance to eliminating humans from the loop in ballistics decision-making, apparently taking a more palatable stance on ``safety'' \cite{Amodei2026}; meanwhile, OpenAI's Sam Altman sought to reassure the public that the U.S. Department of War would adhere to the company's ``safety principles'' when using its tools to wage war \cite{Altman2026}.

``Safety'' always devolves into questions of organizational particulars, institutional arrangements, cultural parameters, and resource constraints \cite{Perrow2011,Downer2024,Vaughan1996,VertesiBoyd2023}. But AI brokers leverage the term's ambiguity to misdirect critics who are concerned about societal impacts. Companies at the heart of the Project of AI suggest that safety is their top priority; they sponsor conferences and technical research into (to put it most starkly) various iterations of the Trolley Problem to assuage stakeholders' concerns. Meanwhile, AI companies race forward, building and deploying models and tools that lack effective guardrails while seeking alignment with dubious values and norms. However, the safety decoy is not a consequence of or fallout from the Project of AI\textemdash it is an integral part of assembling necessary networks of knowledge and capital. As with other decoys, it detracts attention from where the action is while shoring up the networks of power needed to sustain and grow the Project. As the safety decoy helps to reduce public concerns about AI companies, it even creates market opportunities for more or less ``safe'' human-centered tools. All the while, rhetoric of safety bolsters the Project of AI, limiting the very possibility of shelter from the consequences of its adoption.

\subsection{The Regulatory Decoy}

Between the 1990s-2010s, technology industry leaders from Bill Gates to Jeff Bezos asserted in Congressional hearings and antitrust litigation that regulation is the enemy of innovation; meanwhile, critics of the tech industry countered that regulation is the only way to hold the industry accountable. Amid this longstanding polarity, many critics were caught off-guard when key players involved in the Project of AI began clamoring to be regulated. For instance, in May 2023, OpenAI's CEO Sam Altman stood before the U.S. Senate Judiciary Subcommittee on Privacy, Technology, and the Law and asked Congress to create a new government agency that would issue a limited number of licenses to AI model developers, develop safety regulations, and set the standards for AI through the creation and maintenance of shared benchmarks \cite{Kang2023}. The tone of this hearing was unlike previous hearings where tech icons were performatively dressed down by out-of-touch politicians. Instead, Altman told the senators that he was concerned about AI's impact on jobs, its role in geopolitics, and the potential existential risks of rogue AI systems.

Instead of assuming that Altman's pitch for regulation was motivated by a sincere commitment to the law and public authority, we draw attention to how involvement in regulatory (or de-regulatory) processes suits the development of the Project of AI. After decades of eschewing Washington, tech leaders realized that regulation can be strategically beneficial, especially if they have a seat at the regulatory table. Railroad tycoons also famously noted that when captains of industry help craft the laws, those laws can benefit incumbents at the expense of newcomers, protecting the interests of dominant players in industry \cite{Shaffer2009}. Thus a \textit{regulatory decoy} emerges when powerful actors like AI brokers seek regulation to strengthen their structural positions. 

Tech titans have also learned to attend to political pressures. As is true in all major industries, tech companies and their investors often donate to both political parties, as it is useful to stay in politicians' good graces. At the same time, tunes shift when the party in power changes. Hence Altman, who once echoed the Biden Administration's order on ``Safe, Secure, and Trustworthy Development and Use of Artificial Intelligence'' \cite{Biden2023}, later fell in line with President Trump's executive order on ``Removing Barriers To American Leadership In Artificial Intelligence'' \cite{Lima-Strong2025,Trump2025}. Indeed, as legal scholars Julie Cohen and Ari Waldman have noted, U.S. regulatory agencies' approach to governance\textemdash which they dub ``regulatory managerialism''\textemdash can easily be co-opted by corporations \cite{Cohen_Waldman_2023}.

Outside of the U.S., politicians and regulators are more intent on holding the tech industry accountable. Yet they, too, are often enrolled in the project they seek to challenge. Consider the EU's landmark AI Act. In documenting its goals, the EU Parliament stated that it sought to ensure ``safe, transparent, traceable, non-discriminatory and environmentally friendly'' AI, while also highlighting a secondary goal of encouraging ``AI innovation and start-ups in Europe'' \cite{EUParliament2023}. Yet, in regulating AI as a technology, the EU invokes a familiar language of regulation in the public interest, while also embracing a form of techno-legal solutionism \cite{Angel2024} that impossibly assumes technology companies can address or correct complex harms through better product design (if they are forcibly motivated to do so). Meanwhile, the AI Act's weak enforcement mechanisms in many cases entrust firms with assessing the risks posed by their own systems \cite{JonesThorntonDeSilva2025}. In the process, both the technical system and incumbent players become further entrenched.


With the Digital Markets Act (DMA), the EU offers a different path \cite{Hacker_Cordes_Rochon_2024}. Rather than focusing on the products, the DMA targets the network power at the heart of the Project of AI. By labeling key products and companies ``gatekeepers,'' the DMA regulates how networks of actors can operate in relation to one another. Of course, the EU's goal is not to challenge the inevitability of AI but to enable European companies to build an alternative stack in pursuit of a parallel AI Project. Once again, ``for the good guys'' is at the heart of this endeavor. But the DMA's approach focuses on the market-making dimensions of the Project of AI rather than simply on the technology itself.

If AI appears to require urgent regulation in this moment when the postwar power structures are in decline, this is not because its technical capabilities are the cause of instability. Rather, it is because ``AI'' as a rallying cry enables and enacts the ongoing global reconfiguration and infrastructuring of powerful networks in a moment of extreme uncertainty and regulatory disarray. The emerging networks of power that make AI manifest in the world, and their organizing influences\textemdash not the capabilities of an effective chatbot or sloppy video generator\textemdash portends the greatest lasting ``impact'' on ``society.'' Existing work in the social and regulatory sphere must orient to the core of the problem\textemdash financialization, possibilities for the restructuring of companies, and new articulations of monopoly, market-making, and market capture \cite{fourcade2024ordinal,Doctorow_2025,GolkaVanDerZwanVanDerHeide_2024}. If we are to demand that AI systems are held accountable to the relations and (infra)structures of social and public life, we must trouble the conditions of possibility for the construction of this powerful network. Relevant actions could include measures such as increasing tax owed on capital gains, strengthening antitrust enforcement, and closing loopholes that allow investors to hoard wealth.

\section{Discussion: Whither Accountability?}

``AI'' has become an occasion for massive social change, albeit not because a technology is producing unprecedented outcomes. Rather, cash-rich market players are seizing the moment to restructure opportunity and infrastructure around themselves under conditions of global political and economic uncertainty, under the rubric of ``AI.'' As we have seen, one means to shore up these conditions is through mobilizing decoys. It is not the capabilities of an sycophantic chatbot or sloppy video generator that enables the Project of AI to operate as a perceivable phenomenon, but rather the co-construction of a network of power that holds the potential for tremendous and lasting ``impact'' on ``society.'' And if AI appears to be ``changing everything'' at this juncture, this is precisely because ``AI'' provides an ``occasion for structuring'' \cite{barley_technology_1986}, as powerful network brokers pursue ongoing agendas of global reconfiguration and infrastructuring.

Where, then, shall we locate ``transparency'' when ``the algorithm'' is not the most effective, stable, or influential component of this network worth governing? When are we susceptible to stories of inevitability, underhanded maneuvering in regulatory processes, or misjudged ideas of safety? What does it mean to pursue ``accountability'' when a narrow set of elite actors dictate the worlds we should care for through a dizzying array of decoys while controlling the flows of action and possibility? When we attribute the consequences of this moment to which jobs a specific automated technology will re-skill, how LLMs will change education, or how we dial in the pleasantness of an interface, we often leave key actors and activities under-examined. Whether or not AGI appears or robots take our jobs, we will have to contend with long-lasting structural conditions and the vestigial infrastructures from a race among elite corporations and governments to reconstruct the machineries of power in their favor. This moment of AI's cacophonous control over imagined futures will have many afterlives through how AI reifies and makes durable this powerful network of actors. 

When the elements of our critique \textit{underline}\textemdash instead of \textit{undermine}\textemdash the accumulation of power and market-making among a global, hegemonic, and increasingly incontestable and ungovernable elite, we must look elsewhere to push the lever on issues like accountability or fairness. Yet even our attention to morally urgent issues\textemdash such as chatbot-assisted suicide or biased algorithms (which do represent reprehensible harms)\textemdash are enrolled in the Project of AI. This moment therefore calls us to reframe the story of AI governance away from one of technical particulars, and toward the stunning lack of oversight over the growth of the materializing network of power behind AI companies and their products. 

We do not wish to denigrate important work in this community on issues related to AI and society. Our concern is that just when we think we are holding AI companies accountable, we risk embracing a decoy and missing the ultimate site of accountability. We have a responsibility to step back and analyze the entanglements of these systems. Transparency into algorithmic tools is not where accountability lies when the purpose of the game is to build up an incontrovertible governing infrastructure to sustain power among an elite. Fairness in a particular algorithmic outcome or dataset means little in a world in which some of the most massive and enduring inequities since the era of feudalism are sustained through an undemocratic machinery of influence. If key aspects of AI as a phenomenon operate as decoys for technopolitical opportunity, we must instead adopt a new, networked orientation to power and develop the tools necessary to hold the Project of AI to account. 
\subsection{Network Shape-Shifting}


One of the key properties of networked phenomena is the ability to shift agency and accountability by ordering flows elsewhere, displacing or deflecting critiques to other nodes \cite{Hoang2022}. Isolating one node of the network for action (for instance, regulatory oversight or the development of safety tools) allows the rest of the network to re-form out of sight and reassemble itself outside of our purview \cite{ZaidanIbrahim2024}. As we have suggested above, even network brokers are plural and densely interconnected; removing one node from the network does not stop the flow of power or influence. Thus, those who seek inroads for accountability or transparency within a network are instead confronted with its characteristic shape-shifting. This makes the locus for intervention as slippery as the network itself. 

We might assume that CEOs are the right location for intervention as the buck, presumably, stops with them. But the charismatic qualities \cite{Ames2019,Shils1965} that propel the names and faces of these men (and they are all men) to fame are a function of capitalist efforts to make markets \cite{Schumpeter_2008}, not a path to the locus accountability. CEOs deliver smooth, TED-style talks from the mainstage and issue soundbites for the media, but backroom deals among executives, board members, venture capital investment firms, asset managers, politicians, and Wall Street financiers are where the action is. Decision-making is distributed among a networked elite who possess their own mechanisms for consolidating power. Once prospective CEOs plug into hegemonic networks to acquire resources and legitimacy, their agency is severely constrained, as their Board (often made up of representatives from the company's investors) and the Chief Financial Officer (CFO) steer the direction of the company, adjusting its priorities toward return on investment \cite{Shestakofsky2024}.\footnote{ In some cases, dual-class share structures have enabled founders to retain more control over their organizations \cite{Kampmannforthcoming}. Nevertheless, these founders still depend on financial markets to maintain legitimacy, acquire talent, and build products.}

Network-building toward market capture also entails novel forms of organizational and sociotechnical shape-shifting. Companies can easily displace elements like labor, capital, financing, data, and infrastructure to other parts of the network to evade scrutiny \cite{Turner2026}. Supply chains thread these components across the globe, using local conditions to enable resource agglomeration and capital accumulation \cite{CoeYeung2015}. Off-shore ``ghost workers'' \cite{GraySuri2019}, for instance, are essential conscripts in a slippery machinery of circulation. Firms can quickly move data-processing tasks from one location to another, exploiting regulatory gaps or keeping the uncertainty alive long enough to enable market-making while evading oversight \cite{Adams-Prassl_Abraha_Kelly-Lyth_Silberman_Rakshita_2023}. Just as anthropologist Anna Tsing's \cite{Tsing2005} work on flows and frictions in the globalized economy helps us observe networks on the move through conduits that direct activities off-shore or return assets on-shore, this observation inherently destabilizes notions like scale, center, or periphery. Any action, agency, or accountability for the Project of AI can be distributed and re-distributed through the global networked assemblage, not assigned to one person but pulsed across a network like electrical transduction. Inspect one contributing node, and other capacities for action recede out of view. 

Additionally, market-making amid uncertainty requires borrowing relationships, institutions, and ideas from adjacent markets \cite{Fligstein2002,Uzzi1997}. Prior association helps players configure new networks and markets in their favor\textemdash such as by isomorphically arranging their organizations \cite{Caplan-boyd2018,DiMaggioPowell1983}, leveraging interoperability to create or prevent flows \cite{Doctorow_2023}, and setting the price value of various materials and inter-firm exchange \cite{Zelizer1997,UzziLancaster2004,Uzzi1999}. Thus companies straddle multiple institutions and materials that may be subject to oversight in one locale, but evade purview under another configuration. In this, AI companies echo well-established shape-shifting techniques from influential corporations like Facebook, which evaded media industry regulation by arguing that it was not a broadcasting company \cite{NapoliCaplan2017}, or Uber, whose status as a ``transportation network company'' helped it evade existing municipal regulations, exploit workers, and operate in an otherwise protected industry free from oversight \cite{CollierDubalCarter2018,Rosenblat2018,VertesiGoldsteinEnriquezLiuMiller2020}. Even environmental regulation requires characterizing forms of industrial contamination, which companies obfuscate via time-honored techniques of injecting doubt and sustaining ignorance \cite{OreskesConway2010,ProctorSchiebinger2008}.

Finally, the circulation of alternative currencies that secure market-making relations in the Project of AI also confounds oversight \cite{Zelizer1997}. Materials like chips, data center access, and data or model training are not regulated as currencies per se. Yet as they are assetized, their promise and exchange secure networked partnerships among a closed group of elites \cite{BirchMuniesa2020,Uzzi1997,Zelizer1997}, comprising a ``socioeconomic circuit'' \cite{Zelizer1997} for AI. The result is a form of market-making that conscripts digital materials\textemdash proprietary data centers, real estate, the code and conduits that power AI and other properties\textemdash into obligatory passage points \cite{Latour1990} under the control of an elite cadre of network switchers. If prior studies of market-making centered how economic actors contend with states and regulatory agencies, in the current historical moment brokers in the Project of AI enjoy ample opportunity to embed networks of power into the heterogeneous infrastructural assemblages of AI \cite{TanThelen, vanderVlist2024}. Events that interrupt the assembly of such incontrovertible power structures\textemdash such as the release of DeepSeek, OpenClaw, or open-source models\textemdash may offer opportunities to disrupt the networks under construction. On the other hand, because industry leaders can (and often do) take advantage of open-source models in a form of ``co-opetition'' or ``logistical power'' \cite{osborne_characterising_2025,Zhang_Carpano_2023}, we cannot presume that such efforts will undermine the Project of AI. Contestation by external players can be conscripted into the processes of market-making. Indeed, even as new entrants break into the market, they continue to depend upon (and contribute to the profits of) major players like Amazon, Microsoft, and Google, which fuel the industry by providing strategic partnerships, investment capital, software development tools, and computational resources \cite{vanderVlist2024}.

As a network of power, the AI sector relies upon the selective revelation of only certain parts of its heterogeneous network. AI brokers readily redirect flows and shift responsibilities to dodge transparency and accountability. As we have seen with respect to debates about the nature of AI, inevitability discourse, regulatory moves, and even safety procedures, players can and do strategically mobilize or destabilize other parts of their network to evade regulation, oversight, or criticism. When we attend to decoys, then, not only do we do the work of co-constituting the Project of AI; we also allow AI brokers to continue their network-building efforts to maintain their dominance. The result is an industry that appears always to be shape-shifting, chameleon-like, making it hard to pin down where the problems are, let alone to hold the relevant actors accountable. The search for accountability requires new anchoring frames that directly confront network assembly and its slippages.
\subsection{Anchoring Frames for AI Transparency and Accountability}

The future is \textit{not} inevitable. While the Project of AI is extraordinarily powerful, its power is dependent on the continued enrollment of publics and institutions into that project. To that end, it is imperative to recognize when and how we can exercise agency in these fraught moments\textemdash and to resist the distracting decoys nurtured by AI brokers to configure the future on their terms. Reorienting our analyses toward the political economy of AI can help the FAccT community engage with decoys, learn how to understand them \textit{vis-á-vis} wider networks, and take seriously the political power that the Project of AI portends. As we conclude this paper, we offer four possible frames to anchor future work. These both attend to the political economy of AI and open up new ways of imagining AI accountability, fairness, and governance. We hope these anchoring frames can spur new visions for how to realize a more just technologically entangled world.

\textbf{Material sites of network assembly}. The Project of AI requires weaving hefty infrastructures into novel and often unstable alignments, creatively composing relationships and patterns of exchange among heterogeneous actors. More work needs to actively attend to how networks and material infrastructures are prefiguring AI futures. As companies announce novel ``breakthroughs'' or ``features,'' we must interrogate how these moves play into the technopolitical Project of AI by enabling network capture and the acquisition of funding. And we must grapple with how mechanisms of assembly, assetization, and implementation constrain transparency and accountability. 


\textbf{Financing as technopolitical work}. Recent work at the intersection of economic sociology and STS demonstrates how technical work is embedded within financial structures \cite{BirchMuniesa2020,Shestakofsky2024,fourcade2024ordinal,Wajcman_2015}. More scholarship must be attuned to these mechanisms of market making, market capture, and network configuration across the AI sector. Doing so requires attending to the mutual reinforcement of cultural narratives about AI and its circuits of capital. We must also examine how corporate and technical architectures are directed by venture capital and board governance\textemdash and how regulation that focuses on the technology tends to leave the economic and political arrangements driving the Project of AI untouched. Because financial imperatives shape processes of technology development\textemdash and because these decisions are made within a tightly spun network of actors\textemdash we must embrace forms of analysis that bring the social, the technical, and the economic together without determinism \cite{Wajcman_2015}. We may also observe how these governance relations differ across countries and contexts, further contributing to the heterogeneity of AI on the ground \cite{WongChinaAI}.

\textbf{From Objects to Global Flows}. To resist singular definitions or bounded objects with respect to ``AI,'' we suggest switching focus from AI-as-object to the \textit{flows} among sites participating in the global assemblage of AI. Tsing \cite{Tsing2005} identifies flows of capital\textemdash multinational, evasive, opportunistic\textemdash as essential to understanding the construction of market ontologies today. Amid the Project of AI, flows are enabled through shifting relations among individuals, nation-states, corporations, investors, and workers. One locus for transparency and accountability therefore lies with observing shape-shifting and its associated flows, an alternative view of infrastructuring that investigates how connections between nodes are concentrated, fractured, and sustained in particular ways to reproduce relations of power. We might ask: how are flows of data, of people, of capital helped or hindered? Which frictions do they encounter and which are overcome, to whose benefit? To point to a more specific example, how is the project of collecting personal data under surveillance capitalism enrolled unevenly or haphazardly in the Project of AI?



\textbf{Resist Social Solutionism}. Prior work has countered various forms of solutionism, both technological and techno-legal \cite{SelbstFairness2019, cunningham2023grounds, Angel2024}. We must also resist unitary or silo-ed solutionism. Fields naturally have a need to frame the problem of AI in their terms: computing, algorithms, legal regulation, etc. But these framings encourage slipperiness, enable shape-shifting and ontological gerrymandering, and detract from our ability to interrogate the wider Project of AI itself. Focusing too narrowly risks (re)asserting the world-building of AI rather than inviting different cuts or scales that could help to make alternative worlds possible \cite{tsing_mushroom_2015,Haraway_2016,Strathern_1996}. Because no single field will present the ``solution'' to the Project of AI\textemdash and may instead fall prey to decoys\textemdash we must develop transdisciplinary approaches and new knowledge-making practices that follow or interrupt the flows and network shifts upon which the Project of AI relies. A plethora of interventions are needed.
\section{Conclusion}
This paper has sought to sensitize our community\textemdash the FAccT community\textemdash to how AI brokers seek to arrange material, economic, and social networks of power in their favor. Building on a material political economy framework, we have drawn attention to how heterogeneous networks of actors concretize distinct economic and political relations by building them into networked technologies under the auspices of ``AI.'' Political-economic relations among AI brokers form the substrate along which value is ascertained and technological systems like ``AI'' take their shape. Meanwhile, decoys keep critics from engaging with this emerging network of power in a meaningful way. While there are are many ways of observing how ``artifacts have politics'' \cite{winner_artifacts_1986}, our perspective reveals how the Project of AI is one of assembling sociomaterial, networked entities to ground political-economic power plays, constraining and redirecting the path through which resources flow to the benefit of a few players who deflect responsibility and critique. In other words, the game is rigged. 

Addressing the political-economic layer of the Project of AI\textemdash through both analysis and intervention\textemdash is the most pressing, urgent issue associated with the influx of AI. After all, the technical features we encounter are shaped by the interests of those driving this phenomenon. Our community has spent ample time dissecting technical parameters in an earnest attempt to realize a more just and equitable future. But, increasingly, we are being enrolled into the Project of AI by people who have a vested economic and political interest in configuring the future on their terms. We offer this paper as a framework for seeing the forest for the trees, to ensure that we collectively see the Project under assembly before our eyes, to reckon with how our attention to decoys supports the realization of the Project of AI, and to assert alternative loci for transparency, fairness, and accountability.

\section*{Generative AI Statement}

The authors did not use generative AI to write this paper.
\begin{acks}

We are grateful to the networks of scholars who have engaged us, challenged us, and offered us their insightful reflections. We thank the three anonymous reviewers for their generative feedback. We are also deeply grateful to Gil Eyal, Henry Farrell, Marion Fourcade, Tero Karppi, and David Nieborg for comments, to members of Cornell's AI Policy and Practice workshop and Princeton Societal AI for their feedback as we workshopped these ideas, and to Nancy Baym for the location where we first formulated the core argument.  danah boyd thanks the Alfred P. Sloan Foundation's Research Fellowship and Cornell Global Hubs Seed Grant for support. Contributions to this paper and the research behind it by Alex Taylor were made in his capacity as a BRAID Research Fellow and funded by the BRAID programme and the UK’s public funder, AHRC/ UKRI (award AH/X007146/1).

\end{acks}

\bibliographystyle{ACM-Reference-Format}
\bibliography{decoys}
\end{document}